# InSe Schottky diodes based on van der Waals contacts

*Qinghua Zhao, Wanqi Jie, Tao Wang\*, Andres Castellanos-Gomez\*, Riccardo Frisenda\**

Q. Zhao, Prof. W. Jie, Prof. T. Wang
State Key Laboratory of Solidification Processing, Northwestern Polytechnical University, Xi'an, 710072, P. R. China
Key Laboratory of Radiation Detection Materials and Devices, Ministry of Industry and Information Technology, Xi'an, 710072, P. R. China
E-mail: taowang@nwpu.edu.cn

Q. Zhao, Dr. R. Frisenda, Dr. A. Castellanos-Gomez
Materials Science Factory. Instituto de Ciencia de Materiales de Madrid (ICMM-CSIC), Madrid, E-28049, Spain.

E-mail: andres.castellanos@csic.es; riccardo.frisenda@csic.es

**Abstract:** Two-dimensional semiconductors are excellent candidates for next-generation electronics and optoelectronics thanks to their electrical properties and strong light-matter interaction. To fabricate devices with optimal electrical properties, it is crucial to have both high-quality semiconducting crystals and ideal contacts at metal-semiconductor interfaces. Thanks to the mechanical exfoliation of van der Waals crystals, atomically-thin high-quality single-crystals can easily be obtained in a laboratory. However, conventional metal deposition techniques can introduce chemical disorder and metal-induced mid-gap states that induce Fermi level pinning and can degrade the metal-semiconductor interfaces, resulting in poorly performing devices. In this article, we explore the electrical contact characteristics of Au-InSe and graphite-InSe van der Waals contacts, obtained by stacking mechanically exfoliated InSe flakes onto pre-patterned Au or graphite electrodes without the need of lithography or metal deposition. The high quality of the metal-semiconductor interfaces obtained by van der Waals contact allows to fabricate high-quality Schottky diodes based on the Au-InSe Schottky barrier. Our experimental observation indicates that the contact barrier at the graphite-InSe interface is negligible due to the similar electron affinity of InSe and graphite, while the Au-InSe interfaces are dominated by a large Schottky barrier.

## Introduction

Two-dimensional (2D) semiconductors, which hold a great promise for future electronic and optoelectronic applications,[1-3] have motivated a surge of interests of the scientific community. Their ultrathin nature can help in the scaling down of the electronic components to prevent short channel effect,[4] the strong light-matter interaction is interesting for optoelectronic applications and the dangling bond free surface allows stacking various dissimilar functional layers for van de Waals heterostructures building without considering crystal lattice mismatching.[5-8] Thanks to these properties a large amount of outstanding electronic and optoelectronic devices have been fabricated based on 2D metals (graphene/few-layer graphite,[9] metallic transition metal dichalcogenides such as $2H$-$NbS_2$,[10]), insulators (hexagonal boron nitride),[11] semiconductors (semiconducting transition metal chalcogenides such as $2H$-$MoS_2$,[1-2] black phosphorous,[12-13]) and their van der Waals heterostructures. Among the different works reported in literature one can find devices that can work in a broad wavelength range (from ultraviolet to terahertz (THz)),[14] that have ultrahigh photoresponsivity ($\sim 10^{10}$ A/W),[15] fast operating speed, polarization sensitive photodetection or high spatially resolved imaging capabilities.[16-17]

One of the fundamental devices in electronics and optoelectronics is the Schottky diode, which is based on a rectifying metal-semiconductor (M-S) contact.[18-19] Despite their wide use in conventional Si-based electronic systems thanks to their good performances and easiness of fabrication,[20] the amount of studies about Schottky junctions based on 2D semiconductors (as compared to 2D p-n homo- and hetero-junctions[6, 8]) is still very scarce. Until now, the most reported example of 2D-based Schottky diodes is based on the graphene-silicon interface in which the two-dimensional graphene acts as the metal contact and the bulk Si as the semiconductor, thus creating a 2D metal-3D semiconductor junction.[21-24] This kind of device, which is very promising for photovoltaic and broadband photodetection, can be disadvantageous for some applications due to the strong dependency of the Fermi level position in graphene on external factors, such as electric fields, adsorbates or light illumination.[25] The opposite case, which consists of a 3D metal and a 2D semiconductor, is usually realized by evaporating metal contacts onto a 2D semiconductor using conventional metal deposition techniques.[26-29] This fabrication process can introduce defects in the 2D crystal lattice or at the M-S interfaces (in addition to eventual intrinsic defects already present in the 2D material), which can degrade the transport characteristics and induce Fermi level pinning, masking the intrinsic properties of the 2D semiconductors.[19, 30] Moreover, the presence of defects can introduce unwanted hysteresis in the electrical transport and environmental effects.[31-33] Importantly, the lack of dangling bonds in 2D semiconductors allows for the creation of a new kind of electrical contact only mediated by the van der Waals interactions between the metals and the 2D semiconductors by taking advantage of the deterministic transfer method.[5, 34-35] Those van der Waals metal-2D semiconductor interfaces have recently demonstrated to form a reliable electrical contact with low chemical disorder and Fermi level pinning.[30, 36]

Here we study the characteristics of gold (Au) and graphite (Gr) van der Waals electrical contacts on Indium selenide (InSe), an n-type semiconductor, which belongs to the layered III-VIA semiconducting compound group. InSe has gained widespread attention due to its ultrahigh electron mobility (with reported values of ~$10^4$ cm$^2$v$^{-1}$s$^{-1}$ at 4 K and ~4000 cm$^2$v$^{-1}$s$^{-1}$ at room temperature[37]), excellent mechanical properties (the reported Young's modulus of ~20 GPa makes InSe one of the most flexible 2D materials[38]) and super strong light-matter interaction (responsivity up to $10^7$ A/W and detectivity up to $10^{15}$ Jones have been reported [39]).[37-41] We found that the Au-InSe interface is dominated by a large Schottky barrier that we estimate to be approximately 460 meV while the Gr-InSe interface shows a negligible barrier (smaller than 100 meV). We exploit this difference in the barriers to fabricate Schottky diodes based on asymmetrically contacted InSe flakes by van der Waals stacking, which does not require any lithographic process or metallization on 2D semiconductors. The diodes show good performances and follow closely the ideal Schokley equation with a series resistance. Interestingly, our Schottky diodes do not need different metals for the source and drain electrodes, and thus they can be produced using only one metal evaporation step. Our findings demonstrate a new strategy to reach good Schottky diodes by van der Waals stacking based on 2D semiconductors and pave the way for future electronic and optoelectronic applications.

**Results and discussions**

To investigate the properties of thin InSe photodetectors contacted by gold or graphite electrodes we fabricate devices using the deterministic transfer of thin InSe flakes.[34] The InSe flakes were prepared by mechanical exfoliation with Nitto tape (SPV 224) of an InSe single crystal,[38, 42] grown by the Bridgman method.[43] The flakes are then transferred onto a Gel-Film stamp (WF 4x 6.0 mil Gel-Film from Gel-Pak®, Hayward, CA, USA) and identified by optical microscopy. After selecting a suitable flake we transfer it onto pre-patterned gold electrodes on 280 nm SiO$_2$/Si substrate with a deterministic transfer setup described elsewhere.[34] Figure 1a shows the microscope picture of a Au-InSe-Au device fabricated with the above method. The deterministic transfer is also suitable to stack different 2D materials and fabricate van der Waals heterostructures.[5, 35] In a second class of devices we use mechanically exfoliated flakes from natural graphite crystals (HQ Graphene, The Netherlands) to contact an InSe flake on both sides, thereby creating a graphite-InSe-graphite device (Gr-InSe-Gr). Note that we employ graphite (thickness > 12 layers) instead of graphene electrodes to avoid the introduction of a gate dependency and/or a thickness dependency in the work function of the electrodes.[44-45] These Gr-InSe-Gr devices can be fabricated in three transfer steps (see Figure S1 of the Supporting Information for more details about the process). We first select graphite flakes with uniform appearance and having at least a long straight edge. We then transfer one flake contacting only one electrode and then we transfer the second flake contacting only the other electrode. We leave a spatial gap between the two electrodes that can accommodate an InSe flake that can be transferred in the gap contacting the two Gr flakes.

Figure 1b shows a photo of such a Gr-InSe-Gr device, notice that the two gold electrodes contacting the graphite flakes are not shown in the picture. The final devices that we fabricate are based on an InSe flake contacted on one side by a Gr flake and on the other side with a gold electrode (Au-InSe-Gr). The two transfer steps consist in the transfer of a Gr flake contacting only one pre-patterned gold electrode and the subsequent transfer of an InSe flake contacting the Gr flake and the free gold electrode. Figure 1c shows the final picture of a Au-InSe-Gr device. Before performing the optoelectronic characterization of the devices, we leave them in air for 24 hours in order to reach air-stable performances of the devices. In fact, both the photoresponsivity and the operating speed of Au-InSe-Au photodetectors, which are closely related to the presence of trap states in the system, can be tuned by exposing thin InSe to ambient conditions.[42, 46-47] This process, which decreases the trap density in InSe crystals thanks to air passivation, can increase the Au-InSe Schottky barrier by weakening the fermi level pinning at the Au-InSe interface as shown in the case of a Au-InSe-Gr device exposed to air as discussed in Figure S2 of the Supporting Information (see Section S2 of the Supporting Information and our previous work in Ref. [42] for a more detailed discussion). These effects are in agreement with studies reported in the literature about the formation of surface oxides in InSe flakes.[40, 48]

After fabrication we perform the optoelectronic measurements in a home-built probe station kept in high-vacuum ($\approx 10^{-6}$ mbar). Figures 1d-f show the current-voltage (*I-V*) characteristics of the three devices kept in dark (black curves) and under illumination at 530 nm with increasing power density (red curves), all the *I-V*s have been recorded leaving the Si back-gate electrically floating. The inset of each panel shows a schematic of the device under study. In order to compare the characteristics of devices with different geometry we normalize all the *I-V*s by multiplying the current by the dimensionless factor $L/W$, where $L$ is the channel length (calculated from the distance from one electrode edge to the other) and $W$ is the channel width. From the figure one can see that all the three devices show a negligible current in dark (see the Supporting Information Figure S3) and that the current increases with the illumination intensity. The behavior of the *I-V*s for each device is very different and the magnitude of the current is the largest in the Au-InSe-Gr (at positive voltages) and in the Gr-InSe-Gr devices and the smallest in the Au-InSe-Au device.

Starting the discussion from the Au-InSe-Au device *I-V*s, we see that current is symmetric in voltage and has negative curvature, approaching saturation both for large positive and large negative bias voltages (see Figure 1d and Figure S4). These *I-V*s can be explained by the presence of a Schottky barrier at the Au-InSe interface,[49] with a model composed by two Schottky diodes (one for each gold electrode) connected back-to-back with a series resistance, as shown in Figure 1g and in the schematic inset.[50-51] In such a circuit, the voltage drops across the left ($V^L$) and the right ($V^R$) Schottky diodes can be written as:

$$V^L = -\frac{nkT}{q}\ln\left(1 - \frac{I}{I_0^L}\right) \quad \text{and} \quad V^R = \frac{nkT}{q}\ln\left(1 + \frac{I}{I_0^R}\right) \quad (1)$$

where $n$ is the ideality factor of the diodes, $T$ is the absolute temperature, $I$ is the total current flowing through the metal-semiconductor-metal (M-S-M) device and $I_0^L$ ($I_0^R$) is the reverse saturation current of the left (right) diode.[52-54] These saturation currents are related to the Schottky barrier height between gold and InSe $\phi_{B,Au}^0$ according to:

$$I_0^{L/R} = A\,A^*\,T^2\,\exp\left(-\frac{q\,\phi_{B,Au}^0}{kT}\right) \quad (2)$$

where $A$ is the active area (we assume the channel part as the active area in this work) of the device and $A^*$ is the Richardson constant of InSe. The total voltage drop across the the M-S-M device is related to the two voltage drops $V^L$ and $V^R$ and to the series resistance $R_S$ through:

$$V = V^L + V^R + I\,(V)R_S \quad (3)$$

By substituting Eq. 1 in Eq. 3 and adding a current source in series $I_{Photo}$ to take into account the photocurrent generation in the two diodes, the total current $I$ can be written as:

$$I(V) = \frac{I_0^L I_0^R \exp\left[\frac{q(V-I(V)\,R_S)}{nkT}\right] - I_0^L I_0^R}{I_0^L + I_0^R \exp\left[\frac{q(V-I(V)\,R_S)}{nkT}\right]} + I_{Photo} \quad (4)$$

To solve Equation 4 as a function of voltage $V$ we use a numerical method that performs a minimization of the difference between the left and the right side of Equation 4 (The detailed discussion of the fitting model are shown in Figure S5). Figure 1g shows one of the experimental *I-V*s of the Au-InSe-Au device recorded at room temperature ($T$ = 296 K) and with incident power density $P$ = 18 mW/cm², fitted to Eq. 4 using the following parameters: $I_0^L = (6.2 \pm 0.4)$ nA, $I_0^R = (5.1 \pm 0.3)$ nA, $n = (6.2 \pm 1.0)$, $R_S = (60 \pm 10)$ MΩ and $I_{Photo} = (0.03 \pm 0.01)$ nA. From the good quality of the fit we see that the model reproduce well the experimental data indicating that the Au-InSe-Au device has a M-S-M geometry and its behavior is dominated by the Schottky barriers located at the InSe/electrodes interfaces. In general we observe a large value of *n*, larger than 2, which could be due to various sources of non-idealities such as tunneling processes, recombination through mid-gap and traps states in the junction region or image force effects.[24, 53, 55]

Similarly to the Au-InSe-Au device, the *I-V*s measured on the Gr-InSe-Gr device, shown in Figure 1e, are symmetric in voltage reflecting the symmetric electrodes configuration in the device geometry. Notably, in the Gr-InSe-Gr case the curvature of the current versus voltage is positive, meaning that the differential conductance $G(V) = dI/dV$ increases with the voltage and no saturation is observed. Also, the magnitude of the normalized current is approximately one order of magnitude larger in the Gr-InSe-Gr compared to the Au-InSe-Au device in the same conditions. The positive curvature in the *I-V*s and the larger current both indicate a better injection from the Gr electrodes into InSe compared to Au electrodes. By analyzing the *I-V*s of the Gr-InSe-Gr device we found that the current-voltage dependency follows a power law $I \approx V^\beta$. The exponent $\beta$ takes values between 1 and 2,[56] Figure 1h shown one of the *I-V*s fitted with an exponent $\beta = 1.67$ and the inset shows a log-log plot of the same *I-V*. These results indicate that the Gr-InSe interface is not rectifying and that the contact is either ohmic or characterized by a small Schottky barrier $\phi_{B,Gr}^0$.[57-59]

After having determined that the Au-InSe interface shows a rectifying behavior characterized by Schottky barriers $\phi_{B,Au}^0$ we exploited the better injection from the Gr-InSe contact to fabricate Schottky diodes, based on asymmetrically contacted InSe flakes, which present rectifying *I-V*s and photovoltaic characteristics. Figure 1f shows the normalized *I-V*s of such a Au-InSe-Gr device. The shape of the *I-V*s is strongly asymmetric in voltage showing a rectifying behavior and the magnitude of the normalized current for positive voltage is comparable with the one of the Gr-InSe-Gr device. These *I-V*s can be described by the following diode equation:[20]

$$I(V) = I_0 \left\{ \exp\left[\frac{q(V - I(V)R_S)}{nkT}\right] - 1 \right\} - I_{SC} \quad (5)$$

where $I_0$ is the saturation current described by Equation 2. Figure 1g shows one of the experimental *I-V*s of the Au-InSe-Gr device fitted to Equation 5 using the following fitting parameters: $I_0 = (22 \pm 2)$ pA, $n = (2.7 \pm 0.2)$, $R_S = (76 \pm 10)$ MΩ and $I_{SC} = (0.11 \pm 0.02)$ nA. The theoretical curves match perfectly the experimental *I-V* for voltages between -0.5 V and 0.5 V and starts to deviate at larger positive voltages where the experimental current-voltage curve diverges from the linear relation given by $I \approx V/R_S$. Such behavior, which happens after the flat-band condition of the Au-InSe interface, can be described by a dependency of $R_S$ on the inverse of the voltage. The series resistance at the flat-band condition is given by the sum of two components, the sheet resistance of the InSe flake $R_{InSe}$ and the Gr-InSe contact resistance. Assuming that $R_{InSe}$ is constant in this voltage range, the voltage dependency of $R_S$ can then be attributed to the Gr-InSe interface, which is known to have a voltage-dependent density of state and Fermi level or to the onset of space charge limited conduction in graphite as was previously reported.[59-60]

The geometry of our planar InSe Schottky diodes allows us to directly probe the internal electric fields, which should be present at the heterointerfaces, through scanning photocurrent microscopy (SPCM).[61] To acquire SPCM maps we raster scan a light source with wavelength 660 nm focused in a ~1 µm diameter circular spot onto the device surface with the device mounted on a motorized x-y stage, as schematized in Figure 2a. Simultaneously we record the current at a fixed voltage $V_{SD}$ and the intensity of the reflected light. More details about this technique can be found in our previous work.[62] Figure 2b shows the optical picture of the area of the device investigated with SPCM measurements and Figure 2c shows a map of the 660 nm light reflected from that region. All the different features of the device are highlighted by the white dashed contours, used also to align the SPCM maps. Figures 2d-f show the scanning photocurrent maps recorded at three different voltages ($V_{SD}$ = 1 V, 0 V and -1 V). The map recorded at positive voltage, with the diode biased in forward mode (Figure 2d), shows a positive photocurrent (red counts) that is mostly generated at the two contact edges, both at the Au/InSe and at the Gr/InSe interface. Also, a smaller photocurrentis generated in the InSe channel region located between the two electrodes (visible in the in the map as a lighter red shaded region). The rest of the map shows only counts coming from the dark current ($I_{Dark}$ = 0.55 nA at $V_{SD}$ = 1 V). The map recorded at $V_{SD}$ = 0 V (Figure 2e) shows only a region with negative photocurrent centered at the Au-InSe contact edge indicating that an electric field is naturally present at this interface even in absence of an external bias voltage. Notice that the map recorded at zero bias can also be interpreted as a mapping of the short circuit current of the diode. Similar to the previous case, the SPCM map recorded at negative bias (with the diode biased in reverse mode) shows only a region of negative photocurrent generated at the Au-InSe interface with no photocurrent coming from the Gr-InSe interface or from the InSe channel.

Using information from the *I-V*s and the SPCM maps of the Schottky diodes we propose a schematic band structure of the devices under investigation. Figure 3a shows the band structure of the symmetric Au-InSe-Au device at zero external voltage (black curve) and with positive (FV) and negative voltage (RV). The gold electrodes are represented by Fermi distributions with Fermi energy $E_F$ and the InSe by its conductance band minimum. From the *I-V*s of Figure 1d and from the SPCM maps shown in Figure S6 of the Supporting Information one can infer the presence of a Schottky barrier for electrons between n-type InSe and Au. The conductance band minimum of InSe shows a spatial dependence that arises from the charge transfer between metal and semiconductor. Moving from the center of the semiconductor channel toward the electrodes, the band shows an upward bending (in a region generally called depletion layer or space charge region) reaching the highest point at the M-S interface, where the intrinsic electric field is expected to be the largest. The barrier height $\phi_{B,Au}^0 \approx 0.5$ eV, which corresponds to the Schottky barrier height at zero bias, can be calculated according to the Schottky-Mott rule considering the work function of Au $W_{Au} \approx 5.1$ eV and the electron affinity of InSe $\chi_{InSe} \approx (4.55 \div 4.6)$ eV.[24, 29, 63-67] The application of a bias voltage $V_{SD}$ shifts the Fermi energy of each electrode in opposite directions and the voltage drop across each element in the device (left electrode/InSe, InSe channel and right electrode/InSe) depends on the resistance of the particular element (which can in turn depend on the voltage). The width of the depletion layer of the left (right) Schottky barrier is expected to increase (decrease) for RV biasing conditions and to decrease (increase) for FV. Also, the total voltage is expected to drop more across the reversely biased electrode/InSe than the forwardly biased one. In such a device, both at positive and negative voltages, the electrons have to overcome a strong barrier at the electrode reversely biased (the right electrode at positive voltage FV and the left electrode at negative voltage RV). The current is thus limited both for positive and for negative voltages giving rise to the current saturation observed in the *I-V*s at both large positive and negative bias.

Conversely, in the case of graphite electrodes the reported work function values are $\chi_{Gr} \approx (4.5 \div 4.7)$ eV,[68] and thus the expected contact with InSe falls in the range from not rectifying ($\phi_{B,Gr}^0 \approx -0.1$ eV) to slightly rectifying (0.15 eV), Considering the variation in the predicted barrier values, we chose to consider a small Schottky barrier $\phi_{B,Gr}^0 \approx 0.05$ eV noting that a small variation of this value would not change the results of this article. Figure 3b shows the schematic band structure of the Au-InSe-Gr device with three different bias voltages (see also Figure S7 of the Supporting Information for a more detailed version of the band diagram). In this case the asymmetry in the contact barriers with the two different electrodes generates a preferred direction for the current flow. In the forward voltage (FV) mode the Au-InSe contact is forward biased while the Gr-InSe is reversely biased. In such configuration the electrons can flow easily since the only barrier is small and is located at the Gr-InSe reversely biased contact and the total resistance of the device is small and should be limited by the sheet resistance of InSe (the total voltage in this situation drops across the InSe channel and at the InSe/electrode interfaces).[20, 24] In the SPCM map recorded at $V_{SD}$ = 1 V shown in Figure 2d the observation of a positive photocurrent located at both the Au-InSe and the Gr-InSe interfaces is consistent with the band diagram of Figure 3b in FV voltage configuration (see also Figure S7 of the Supporting Information). On the other hand, in the reverse voltage (RV) biasing the strong Schottky barrier $\phi_{B,Au}^0$ is reversely biased and thus only a small current can flow through the device, which is characterized by a large resistance dominated by the Au-InSe contact resistance (with the voltage dropping mostly across this interface). In order to estimate the barrier height $\phi_{B,Au}^0$ we studied the temperature dependency *I-V*s of the device kept in dark, reported in Figure 3c. The *I-V*s recorded while increasing the temperature from room temperature (*T* = 297 K) to *T* = 452 K show a current increase both in the forward bias and in reverse bias and a decrease of the rectification. We fit all these *I-V*s using Eq. 5 and from the results we use the saturation current $I_0$ and the temperature to make Richardson plot of $\ln(I_0/T^2)$ vs. $1000/T$ shown in Figure 3d. By rewriting Eq. 2 as:

$$\ln(I_0/T^2) = \ln(A\,A^*) - \frac{q\,\phi_{B,Au}^0}{kT} \tag{6}$$

one can see that the slope in the Richardson plot gives the barrier height $\phi_{B,Au}^0$. From the fit shown in Figure 3d we extract a barrier height of $\phi_{B,Au}^0 = (0.46 \pm 0.06)$ eV, which is comparable to the value predicted by the Schottky-Mott rule.[20, 63-64]

After having established the energetic of the devices under study we characterize the optoelectronic performances of the Schottky diode. Figure 4a shows the current recorded as function of time *I-t*s with the device kept under a square wave modulated illumination at 530 nm. At the beginning of the traces, when the device is kept in dark, the current flowing through it is on the order of few pA both in FV and in RV bias ($V_{SD}$ = 2 V and $V_{SD}$ = -2 V respectively). After switching on the illumination we observe a positive photocurrent at $V_{SD}$ =

2 V and a negative photocurrent at $V_{SD}$ = -2 V. The photodiode geometry allows detecting an external illumination also in absence of an applied voltage as can be seen in the *I-t* recorded at zero bias that shows a negative photocurrent under illumination. From the plot it is evident that both the magnitude of the photocurrent and the response time change dramatically under FV and RV biasing. To better appreciate this fact, we plot in Figure 4b the previous *I-t* in a semi-logarithmic representation, plotting the current in absolute value. The dark current recorded at the beginning of the trace, at zero bias and at -2 V, is on the order of 1 pA (comparable to the noise floor of the setup), while at 2 V is about two orders of magnitude larger. After switching on the illumination, the photocurrent with the device in FV biasing grows to of approximately 3 nA, which is about one order of magnitude larger than the current generated at zero bias and in RV both showing a similar photocurrent of 300 and 400 pA respectively.

The strong bias dependency of the photocurrent can be seen in Figure 4c where we collect the total current flowing through the device in dark and under illumination extracted from *I-t*s recorded at different bias in the range -5 V to 5 V. At negative voltages the photocurrent shows only a weak dependency on the applied voltage, going from -0.42 nA at -5 V to -0.30 nA at 0 V corresponding to an increase rate of ~0.03 nA/V. At positive voltages the rise is much faster and we extract a rate of ~2 nA/V, approximately two orders of magnitude larger than the reverse biased regime. This large change in the magnitude of the photocurrent is accompanied by a change in the time response of the device. We define the rising time $\tau_r$ and the decay time $\tau_d$ using the 10%-90% rule,[17] in which we extract the time that takes for the signal to go from 10% of the saturation to 90% of this value, as shown in Figure 4a in the case of $\tau_r$. Already inspecting the traces in Figures 4a and b one can see that both rising and decay times are much faster when the device is operated at zero bias and at -2 V compared to 2 V. This fact is clearly visualized in Figure 4d where the extracted rising and decay times are plotted as a function of voltage. For negative voltages the extracted rising times of 40 ms are probably upper bounds limited by the time resolution of the measurement setup while the decay times are on the order of 280 ms, contrarily when a positive voltage is applied the rising times increase to more than 13 s and the decay time can reach 1.8 s, showing a change in response time of more than one order of magnitude when passing from FV to RV biasing.

Finally we characterize the illumination power dependency of the device. Figure 5a shows a set of *I-V*s plotted in semi-logarithmic scale and recorded at room temperature with the device kept in dark and under illumination at 530 nm with power densities going from 1.2 mW/cm² to 18 mW/cm². To fit the *I-V*s we use equation 5 and the results are shown in Figure 5a and b. As can be seen from Figure 5a the agreement between the *I-V*s predicted by the Schottky model and the experimental *I-V*s is excellent, which can reproduce well the increase in both the saturation current in reverse biasing and in the forward voltage current. In Figure 5b we show the dependency of the Schottky diode ideal factor, series resistance and reverse saturation current on the incident optical power. When increasing the illumination power the ideal factor $n$, which is related to the recombination mechanism of the free charge carriers in the device, increases from 2.2 to 2.7, at the same time the series resistance decreases of approximately one order of magnitude, going from 500 MΩ to 70 MΩ and the reverse saturation current increases linearly with the illumination power going from 2 pA to 20 pA. The linear increase in $I_0$ can be explained by the lowering of the Schottky barrier $\phi_{B,Au}^0$ under illumination,[69-70] also confirmed by the lower value of the Schottky barrier estimated using the temperature dependency of the *I-V*s under illumination shown in the Supporting Information (see Figure S8). The increase in the ideality factor is consistent with the lower barrier and indicates the activation of additional non-ideal recombination paths under high light intensity and large number of photogenerated charge carriers. The higher illumination power also induces a decrease of $R_s$, which can be understood from an increase of the carrier density in InSe. Writing the current flowing in a semiconductor assuming only drift processes, $I = (N + P)\, q\, v$,[20] where $(N + P)$ is the density of free carriers and $v$ is the drift velocity, one can see that the resistance, which is proportional to $(N + P)^{-1}$, is expected to be inversely proportional to the optical power $P_{Opt}$. The middle panel of Figure 5b shows the excellent agreement between the experimental $R_s$ values and the theoretical function $R_s (P_{Opt}) = a \cdot P_{Opt}^{-1} + b$ with $a$ and $b$ as fitting parameters equal to respectively 570 MΩ·cm²/mW and 55 MΩ.

Apart from the fitting parameters defined by the diode current-voltage equation, two important quantities especially relevant in solar cells are the open-circuit voltage $V_{OC}$ and short-circuit current $I_{SC}$. Figures 5d and 5e show the power dependency of these two quantities. As can be seen from the good agreement between the fit and the experimental plots, $V_{OC}$ depends logarithmically on the incident power and $I_{SC}$ depends linearly on the power. These two dependencies confirm that the photocurrent is predominantly caused by the photovoltaic effect as opposed to thermoelectric effects. The linearity of the short-circuit current is also evident in the log-log plot of the photocurrent generated at zero bias voltage versus illumination power, shown in Figure 5f. From a power law fit to the function $I \sim P_{Opt}^{\alpha}$,[17] using the exponent $\alpha$ as the only fitting parameters, we find that at 0 V $\alpha$ = 0.94 and that in reverse bias at -1 V $\alpha$ = 1. Contrarily, at 1 V in the forward voltage regime the photocurrent shows a sublinear dependency on the incident power with $\alpha$ = 0.66. The voltage dependency of $\alpha$ can be explained by the different photocurrent generation mechanism in reverse and forward bias.[17, 42] In the reverse bias (and zero bias) cases, the presence of an (intrinsic) electric field in the device promotes the formation of a depletion region that causes the photodetector to operate as a normal photodiode with fast operation speed, but limited responsivity. In this case, the photogenerated electron-hole pairs in InSe can be efficiently separated by the electric field present at the Au-InSe interface and the minority carriers can quickly drift away from the InSe and can be collected into the Au electrode. This fast removal of the photogenerated minority carriers from the InSe channel prevents the trapping of these carriers and the subsequent photogain mechanism that can result in lower response speed. For forward bias the external electric field is applied in the opposite direction in respect to the intrinsic one, thus reducing the depletion layer width. In absence of such a space charge region, the minority carriers can get trapped by holes trapping centers. This trapping processes effectively increase the lifetime of the majority carriers and introduce photogain in the system, which can increase the responsivity at the expenses of the response speed. The value close to 1 assumed by $\alpha$ in the reverse biasing and zero bias conditions confirms that hole trapping centers are not effective in these regimes, corresponding to a linear dependency of the photocurrent on the illumination power (and a power independent responsivity). On the other hand, the participation of traps in the FV regime reduces the value of $\alpha$ to 0.66, which indicates that the rate of carriers trapping increases with the illumination power and thus at high injection level fewer carriers participate to the total current as compared to low injection. This is consistent with the voltage dependency of the response time and responsivity of the device which shows slower photodetection with higher responsivity in the FV biasing and faster photodetection but with lower responsivity in the RV biasing.

**Conclusion**

In summary, we investigated different devices based on thin InSe with different van der Waals electrical contacts. In the case of Au electrodes we find a rectifying Schottky contact with a sizeable barrier between InSe and Au, estimated to be ∼460 meV by temperature dependent measurements a value that is consistent with Schottky-Mott rule, while in the case of graphite the contact gives rise to a negligible contact barrier (smaller than 100 meV) thanks to the similar electron affinity of thin InSe and graphite. Exploiting this strong contact barrier difference, we can easily fabricate a Au-InSe-Gr Schottky diode, based on symmetric pre-patterned gold electrodes, with asymmetric van der Waals contacts. The asymmetric contacted device shows *I-V* curves that follow perfectly the Schottky diode equation in a large bias range. Furthermore, also the optoelectronic measurement of the devices is in agreement with the predicted Schottky diodes behaviors under reverse and forward biasing conditions. Our results show the reliability of electronic and optoelectronic properties of van der Waals Schottky contacts. The easiness of fabrication, which comes from the van der Waals stacking by dry deterministic transfer, and the reproducible Schottky barrier formation are important factors for future applications of 2D semiconductors-metal systems.

**Materials and Methods**

**Sample fabrication.** The mechanical exfoliation method with scotch tape and Nitto tape (Nitto Denko® SPV 224) was used for thin graphite and InSe flakes fabrication onto a Gel-Film stamp (WF 4x 6.0 mil Gel-Film from Gel-Pak®, Hayward, CA, USA). After optical microscopy inspection under transmission mode (Motic® BA310 MET-T), the selected graphite flakes with uniform thickness were deterministically transferred from the Gel-Film onto the pre-patterned Au (50nm)/SiO$_2$ (280nm)/Si substrates (Osilla®) to fabricate Gr-Gr or Gr-Au electrodes. Then the selected InSe flakes (with thicknesses in the range 20-25 nm) were deterministically transferred, bridging the gaps between the different electrodes to fabricate Au-InSe-Au, Gr-InSe-Gr and Au-InSe-Gr devices. The channel length of pre-patterned gold electrodes is 30 μm and the van der Waals heterostructures fabrication processes are present in the Figure S1 in Supporting Information.

**Optoelectronic characterizations.** All the electrical and optoelectronic characterizations of devices were carried out in a homebuilt high-vacuum (~$10^{-6}$ mbar, room temperature $T$ = 296 K - 452 K) chamber. A source-meter source-measure unit (Keithley® 2450) was used for performing the electrical measurements ($I$-$V$, $I$-$t$). The light source with wavelength 530 nm is provided by light emitting diodes (LEDD1B – T-Cube LED driver, Thorlabs®), projected onto the sample surface by a zoom lens by coupling to a multimode optical fiber at the LED source, and a light spot on the sample with the diameter of 600 μm can be observed. The scanning photocurrent maps under various source-drain voltage bias (-1 V, 0 V and 1 V) are acquired with an home-made scanning photocurrent system based on a modified microscope (BA310Met-H Trinocular, MOTIC), a motorized XY scanning stage (8MTF, Standa) and a source-meter source-measure unit (Keithley® 2450). In this setup, the light coming from a 660 nm fiber coupled laser is focused in a spot with a diameter of ~1 μm onto the sample surface with a 20X miscoscope objective (total power 0.7 μW), a more detailed description can be found in our previous work.[62]

**AFM measurements.** Thin InSe flakes thickness measurement was carried out by an ezAFM (by Nanomagnetics) atomic force microscope operated in dynamic mode. The force constant of the cantilever used is 40 Nm$^{-1}$ and the resonance frequency 300 kHz (Tap190Al-G by Budget Sensors).

**ACKNOWLEDGEMENTS**

This project has received funding from the European Research Council (ERC) under the European Union's Horizon 2020 research and innovation programme (grant agreement n° 755655, ERC-StG 2017 project 2D-TOPSENSE). EU Graphene Flagship funding (Grant Graphene Core 2, 785219) is acknowledged. RF acknowledges support from the Spanish Ministry of Economy, Industry and Competitiveness through a Juan de la Cierva-formación fellowship (2017 FJCI-2017-32919). QHZ acknowledges the grant from China Scholarship Council (CSC) under No. 201700290035. TW acknowledges support from the National Natural Science Foundation of China: 51672216.

**COMPETING INTERESTS**

The authors declare no competing financial interests.

**FUNDING**

Spanish Ministry of Economy, Industry and Competitiveness: Juan de la Cierva-formación fellowship 2017 FJCI-2017-32919

EU H2020 European Research Council (ERC): ERC-StG 2017 755655

EU Graphene Flagship: Grant Graphene Core 2, 785219

National Natural Science Foundation of China: 51672216

**Supporting Information**

Fabrication and characterization of pristine devices, statistics of different devices, theoretical *I-V*s of the back-to-back diode model and scanning photocurrent microscopy measurements of a Au-InSe-Au device, estimation of the Au-InSe Schottky barrier height under illumination.

**FIGURES**

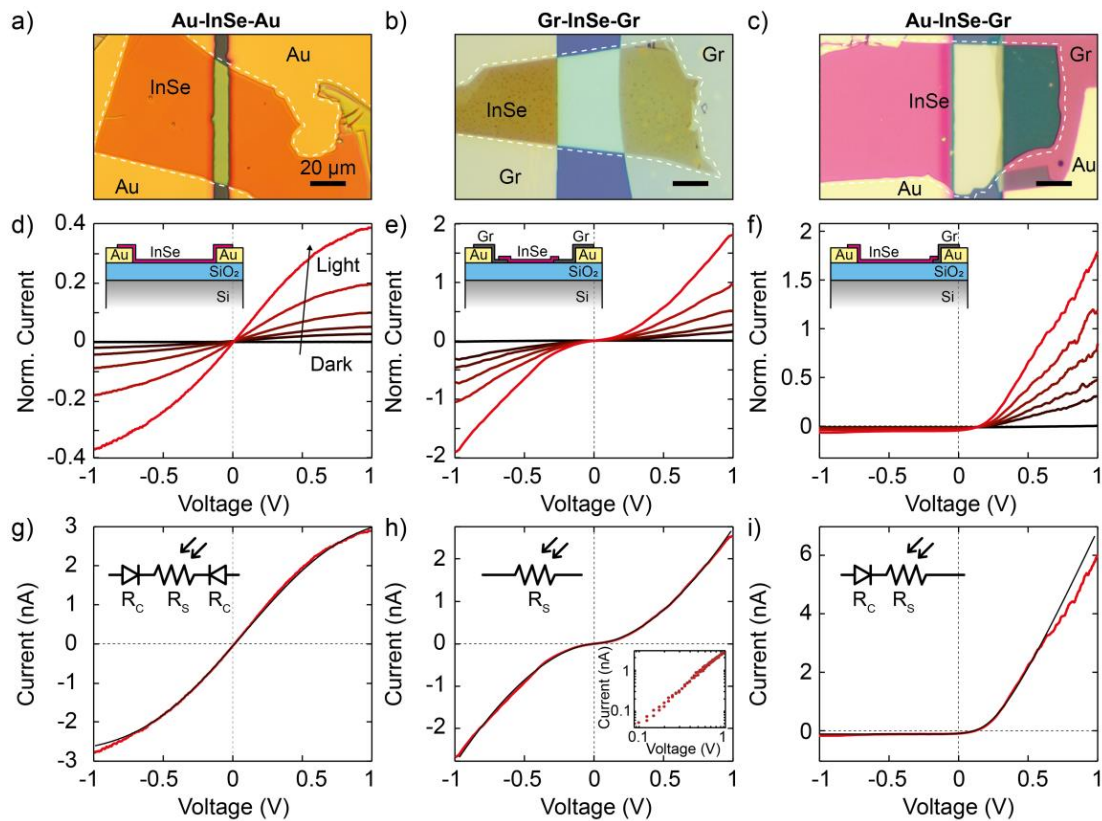

**Figure 1**: a-c) Optical pictures of symmetric and asymmetric devices based on few-layer InSe, gold electrodes (Au) and graphite (Gr). d-f) Current-voltage characteristics of the three devices shown in panels (a)-(c) in dark and under 530 nm illumination with different intensity (from 1.2 mW/cm$^2$ to 18 mW/cm$^2$). g-i) Experimental *I-V* curves recorded with the largest illumination density (red curve) and theoretical fit (black curve). The *I-V* of Au-InSe-Au is adjusted to the back-to-back Schottky diode model, the one of Gr-InSe-Gr to a power law with exponent between 1 and 2 and the *I-V* of Au-InSe-Gr is fitted to a Schottky diode with series resistance model.

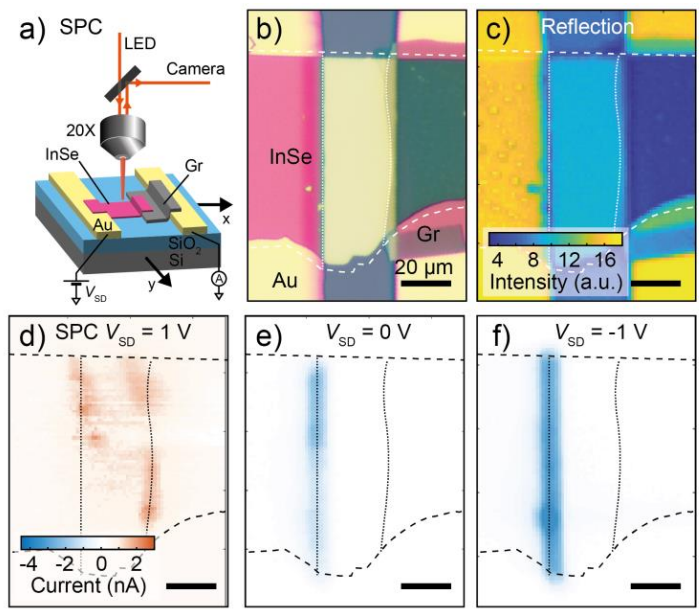

**Figure 2**: a) Schematic drawing of the scanning photocurrent (SPC) technique. b) Optical picture of the area of the Au-InSe-Gr studied with SPC. c) Laser reflection map recorded with 660 nm light spot at the same time of the photocurrent. d-f) Photocurrent maps of the device recorded in three different biasing conditions: forward voltage (1 V) (d), zero bias (e) and reverse voltage (-1 V) (f).

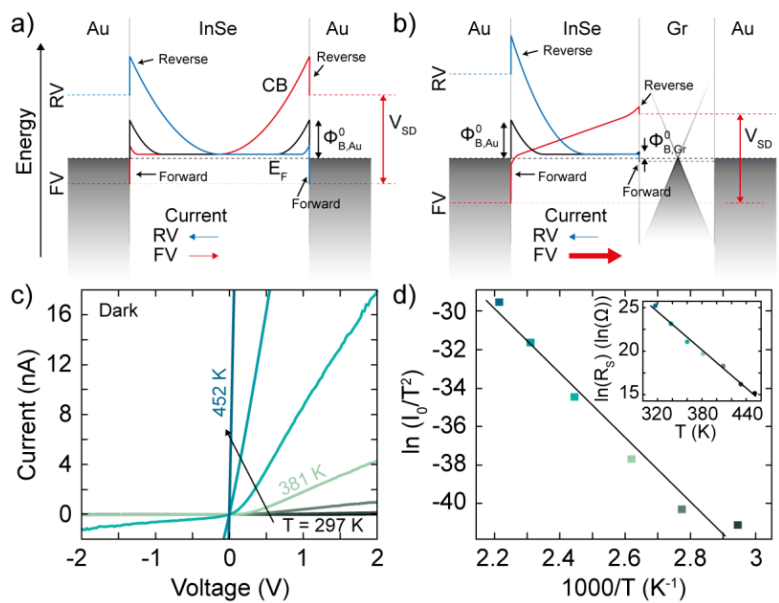

**Figure 3**: a-b) Band diagram of a Au-InSe-Au (a) and Au-InSe-Gr (b) device at zero voltage (black curve), at positive voltage (red curve, FV) and at negative voltage (blue curve, RV). The dashed line shows the Fermi level energy ($E_F$) and there are indicated the Schottky barriers of InSe with gold $\Phi^0_{B,Au}$ and with graphite $\Phi^0_{B,Gr}$. c) Current-voltage characteristics of Au-InSe-Gr recorded at different temperatures in dark conditions. d) Richardson plot of the saturation current of Au-InSe-Gr extracted from fits to the *I-V*s in panel (c). Inset: natural logarithm of the series resistance ($R_S$) extracted as a function of temperature.

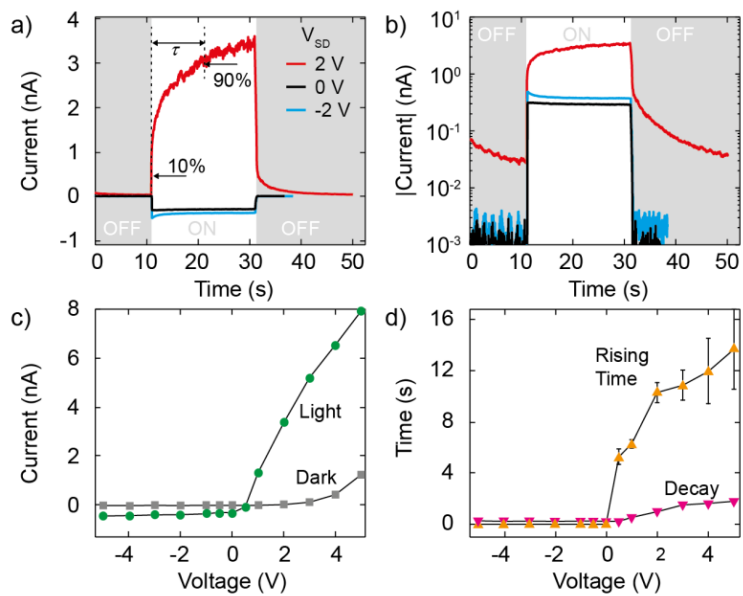

**Figure 4**: a-b) Current in linear scale (a) and current absolute value in logarithm scale (b) at different $V_{SD}$ (-2 V, 0 V, 2V) recorded in an Au-InSe-Gr Schottky diode as a function of time (*I-t* curves) while turning on and off a 530 nm light source with a power density of 33.4 mW/cm². c) Light (green) and dark (gray) current extracted from a series of *I-t* curves as a function of bias from -5 V to 5 V while the 530 nm light source under on and off state. d) Rising time (yellow) and decay time (purple) as a function of bias from -5 V to 5 V extracted from *I-t* curves.

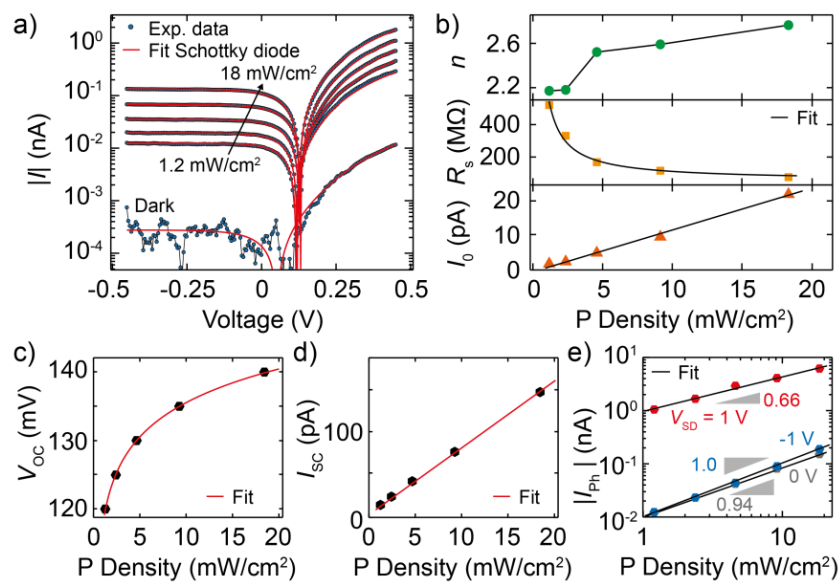

**Figure 5**: a) Experimental (blue dots) and fitted (red curves) dark and under various 530 nm illumination powers *I-V* curves plotted in semi-logarithmic scale based on current absolute value. b-d) The fitted parameters $I_0$, series resistance $R_s$, ideal factor $n$ (b), open circuit voltage ($V_{oc}$) (c) and short circuit current $I_{sc}$ (d) as a function of 530 nm illumination power density extracted from panel a. f) Experimental photocurrent as a function of illumination power at 530 nm recorded at three different bias (-1 V, 0 V, 1 V).

# Supporting Information:

# InSe Schottky diodes based on van der Waals contacts

*Qinghua Zhao, Wanqi Jie, Tao Wang\*, Andres Castellanos-Gomez\*, Riccardo Frisenda\**

Q. Zhao, Prof. W. Jie, Prof. T. Wang
State Key Laboratory of Solidification Processing, Northwestern Polytechnical University, Xi'an, 710072, P. R. China
Key Laboratory of Radiation Detection Materials and Devices, Ministry of Industry and Information Technology, Xi'an, 710072, P. R. China
E-mail: taowang@nwpu.edu.cn

Q. Zhao, Dr. R. Frisenda, Dr. A. Castellanos-Gomez
Materials Science Factory. Instituto de Ciencia de Materiales de Madrid (ICMM-CSIC), Madrid, E-28049, Spain.

E-mail: andres.castellanos@csic.es; riccardo.frisenda@csic.es

### Section S1 - Device fabrication

Figure S1 shows microscope optical pictures acquired during the fabrication of a Gr-InSe-Gr device (S1a) and of a Au-InSe-Gr Schottky diode (S1b).

Figure S2 shows the optoelectronic characterizations of a Au-InSe-Gr device performed just after the fabrication in its pristine state and after being exposed to air. Figure S2a shows the current-voltage characteristics of the device kept in dark in pristine conditions (black curve) and after 13 hours in air (red curve) shown with semi-logarithmic scale. The current decreases around three orders of magnitude after the exposition to air and the current observed at negative voltages become comparable to the noise level of our measurement. Figure S2b shows similar measurements performed while keeping the device under external illumination (wavelength 530 nm). Also in this case a clear decrease of the total current can be observed in the aged device compared to the pristine state. Interestingly the device after 13 hours of exposure to air develops an open circuit voltage ($V_{OC}$) that is absent in the pristine state. The decrease in photocurrent as a function of exposure to air can be better observed in Figure S2c. Figure S2d shows the increase in the open circuit voltage, which in the pristine device is 0 V and reaches 0.13 V after 13 hours in air. Both the decrease in photocurrent and the increase in the open circuit voltage can be explained by an increase in the height of the Schottky barrier formed at the Au-InSe interface. Figure S2e-f shows the power dependency of the photocurrent at negative (S2e) and positive (S2f) voltages for the pristine device and after 13 hours in air. In all cases, the data appear to follow a linear trend in the log-log graphs that can be mathematically described by a power law $I_{ph} = P^{\alpha}$, where P is the optical power and the exponent α is the fitting parameter. Interestingly, in the case of the pristine device the power law exponent has a value of 0.34 ± 0.01 both for negative and positive voltages, while the aged device shows an asymmetric α, which is equal to 1.02 ± 0.03 at negative voltages and 0.66 ± 0.02 for positive voltages. The low value of α in the pristine state can be explained by the presence of traps in the system and negligible Schottky barrier, which also explain the large photoresponsivity observed in this state. In the aged state, the two different values of α for positive and negative voltages indicate the presence of a sizeable Schottky barrier, which induce an intrinsic electric field in the device.

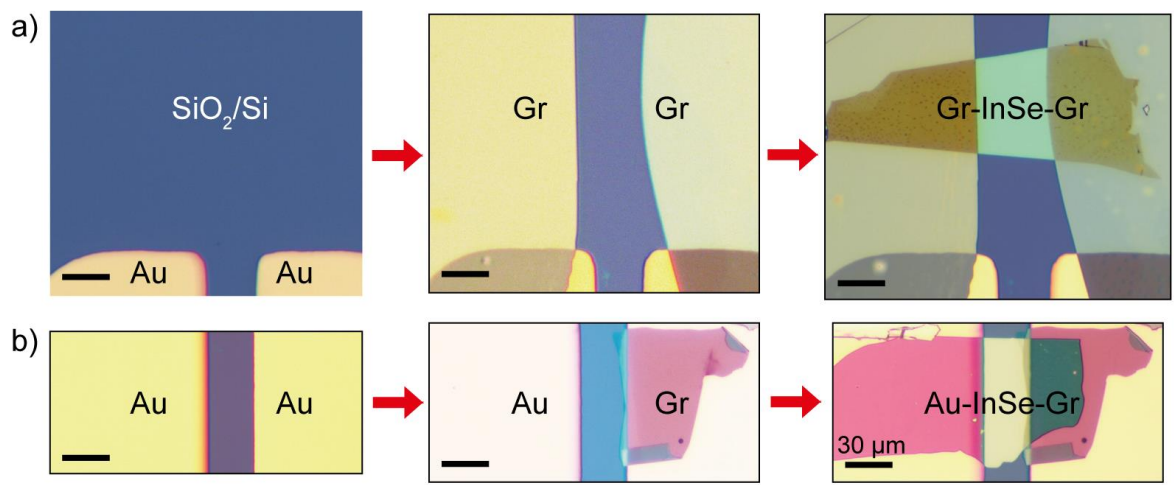

**Figure S1:** Devices with the geometries of Gr-InSe-Gr (a) and Au-InSe-Gr (b) fabrication processes.

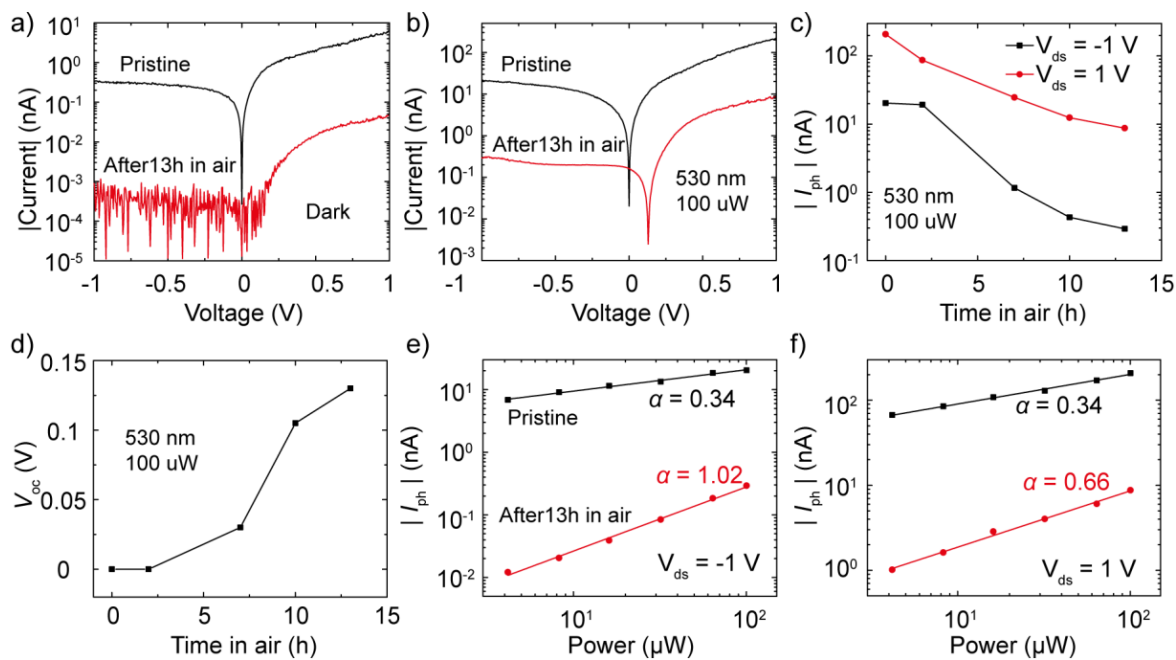

**Figure S2:** Performance evolution of a Au-InSe-Gr Schottky diode as a function of air-exposure time. (a), (b) The *I-V* curves plotted in semi-logarithmic scale recorded in dark conditions (a) and under illumination with 530 nm light source (b) with the states of pristine (black line) and after 13 hours air exposure (red line). (c), (d) The variation of photocurrent at 1V (red) and -1 V (black) (c) and open circuit voltage ($V_{oc}$) extracted from *I-V* curves recorded as a function of air exposure time. (e),

(f) The power dependent photocurrent recorded at -1 V (e) and 1 V (f) of the Au-InSe-Gr device in pristine state and after 13 hours air exposure.

## Section S2 - Devices statistics

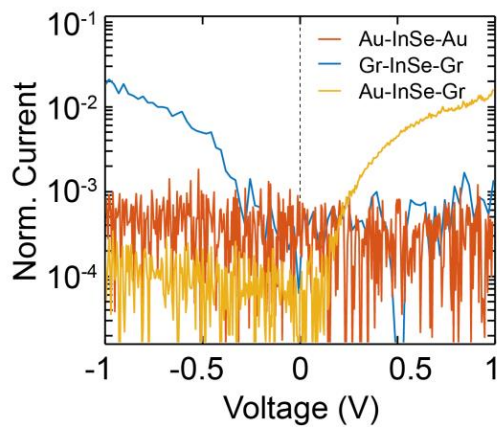

**Figure S3:** Semi-logarithmic representation of the experimental *I-V*s of the three devices discussed in Figure 1 of the main text recorded in dark.

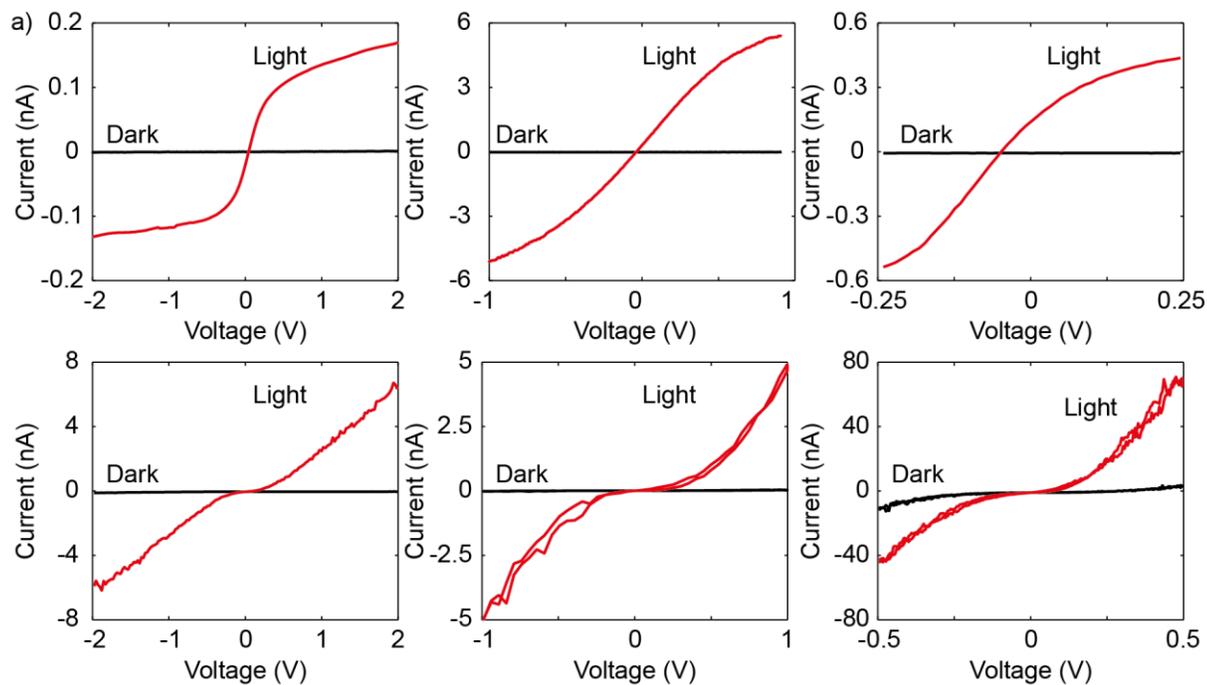

**Figure S4:** Experimental *I-V*s of different symmetrically contacted InSe devices: Au-InSe-Au back-to-back Schottky diodes devices (top row) and Gr-InSe-Gr (bottom row)

## Section S3 – Back-to-back Schottky diode current model

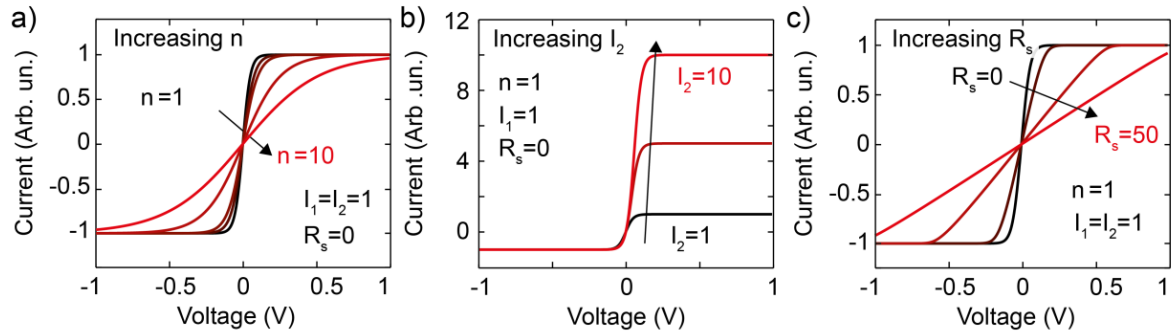

**Figure S5:** a-c) Theoretical *I-V*s calculated from the back-to-back diode model using Equation 4 of the main text for different parameters values. In each panel we vary one of the parameters of the model ($n$, $I_1$, $I_2$ or $R_S$) while keeping the other parameters fixed to show the influence of each parameter on the *I-V* curves predicted by the back-to-back diode model.

## Section S4 – Band diagrams of the Schottky-barrier devices

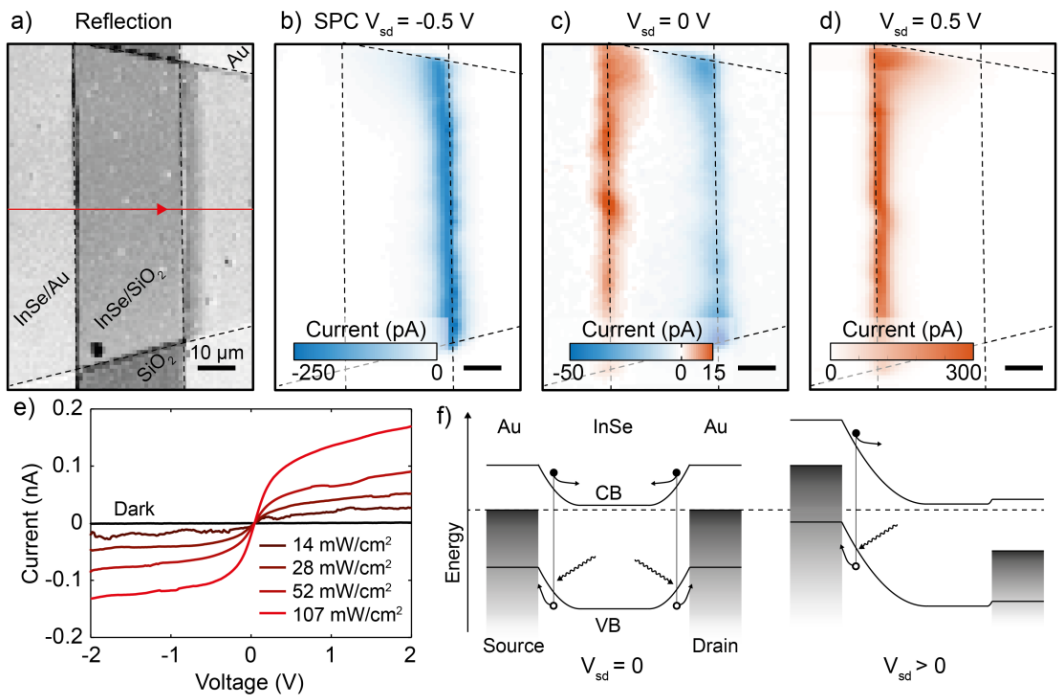

**Figure S6:** a) Laser reflection map recorded with 660 nm light spot at the same time of the photocurrent of a Au-InSe-Au device. b-d) Scanning photocurrent maps of the device recorded for three different bias voltages. e) Current-voltage characteristics of the device in dark and under global illumination at 660 nm with increasing power density. f) Schematics of the band structure of the device with and without a bias voltage applied. The wiggly lines represent photons (with energy larger than the InSe bandgap) hitting the device and generating a photocurrent.

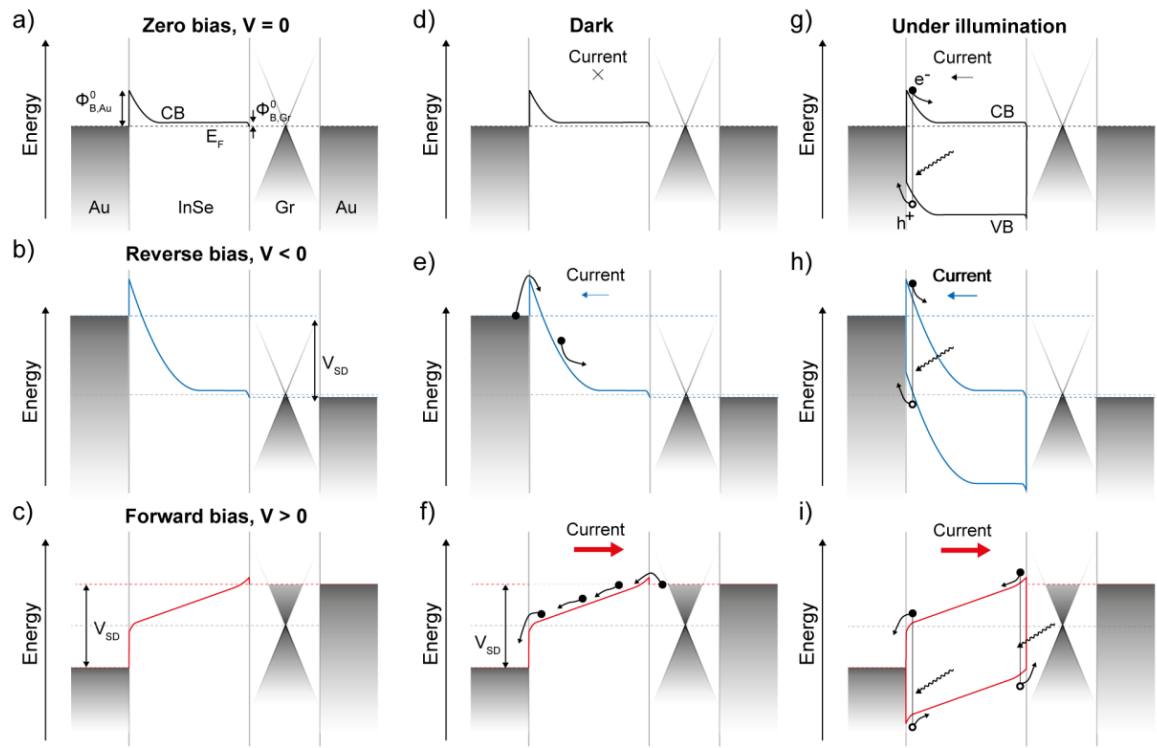

**Figure S7:** a-c) Band diagram of an Au-InSe-Gr device at zero voltage (black curve), at reverse or negative voltage (blue curve) and at positive or forward voltage (blue curve). The dashed line shows the Fermi level energy ($E_F$) and there are indicated the Schottky barriers of InSe with gold $\Phi^0_{B,Au}$ and with graphite $\Phi^0_{B,Gr}$. d-f) Same diagrams of panels (a-c) with indicated the electron injection and the current flow in dark conditions. g-i) Same as panels (d-f) but under illumination conditions.

### Section S5 – Estimation of the Au-InSe Schottky barriers under external illumination

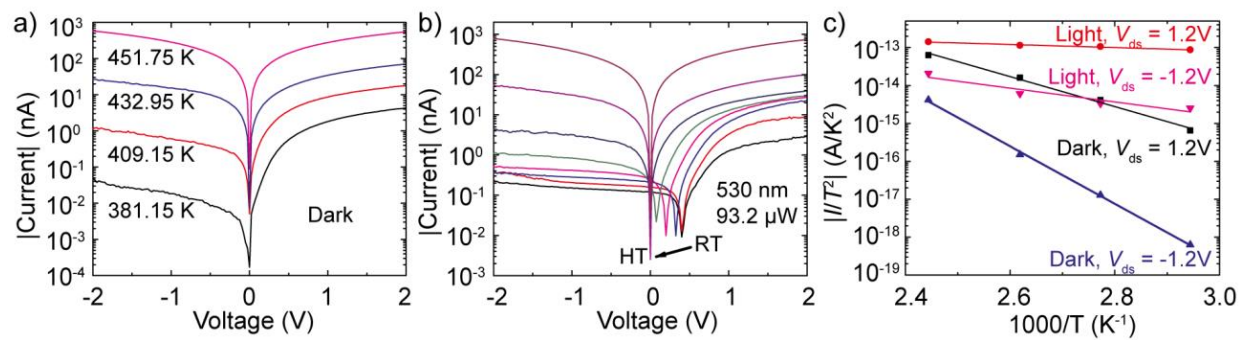

**Figure S8:** Temperature dependent *I-V* curves recorded in dark (a) and under 530 nm illumination with the power of 93.2 µW (b) with a high condition of ~$10^{-6}$ mbar. c) The Schottky barrier extraction at the bias of 1.2 V and -1.2 V in dark and light conditions, clear observation is that the Schottky

barriers ($\Phi_{B, light, 1.2 V}$ = 35.2 meV, $\Phi_{B, light, -1.2 V}$ = 163.8 meV) under illumination both at 1.2 V and -1.2 V bais smaller than them in dark ($\Phi_{B, dark, 1.2 V}$ = 342.9 meV, $\Phi_{B, light, -1.2 V}$ = 657.1 meV).

## TOC Graphics:

## InSe Schottky diodes based on van der Waals contacts

*Qinghua Zhao, Wanqi Jie, Tao Wang\*, Andres Castellanos-Gomez\*, Riccardo Frisenda\**

Q. Zhao, Prof. W. Jie, Prof. T. Wang
State Key Laboratory of Solidification Processing, Northwestern Polytechnical University, Xi'an, 710072, P. R. China
Key Laboratory of Radiation Detection Materials and Devices, Ministry of Industry and Information Technology, Xi'an, 710072, P. R. China
E-mail: taowang@nwpu.edu.cn

Q. Zhao, Dr. R. Frisenda, Dr. A. Castellanos-Gomez
Materials Science Factory. Instituto de Ciencia de Materiales de Madrid (ICMM-CSIC), Madrid, E-28049, Spain.

E-mail: andres.castellanos@csic.es; riccardo.frisenda@csic.es

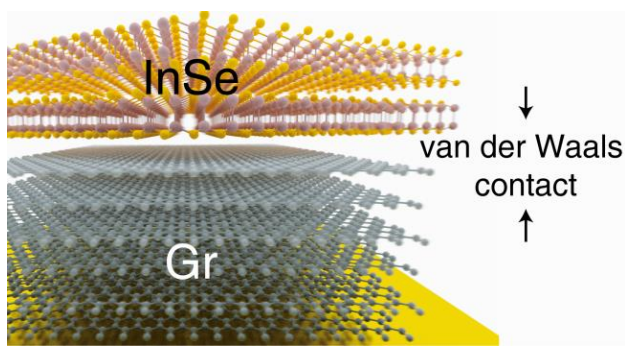